\documentclass[aps,prl,twocolumn,groupedaddress,showpacs]{revtex4}
\usepackage{graphics}
\usepackage{amsmath}
\usepackage{epsfig}
\usepackage{epsf}

\begin{document}
\title{Partonic calculation of $\gamma Z$ Exchange Corrections to
\\Parity-Violating Elastic Electron-Proton Scattering}

\date{\today}

\author{Yu Chun Chen$^{1}$,
Chung Wen Kao$^2$, and Marc Vanderhaeghen$^{3}$ \\
$^1$ Institute of Physics, Academia Sinica, Taipei, Taiwan\\
$^2$Department of Physics, Chung-Yuan Christian University,
Chung-Li 32023, Taiwan\\
$^3$ Institut f\"ur Kernphysik, Johannes Gutenberg-Universit\"at,
Mainz, Germany\\
}
\begin{abstract}
We calculate $\gamma Z$ exchange
corrections to the parity-violating asymmetry of the elastic
electron-proton scattering in a parton model using the formalism
of generalized parton distributions(GPDs). We also examine the validity of the zero-momentum-transfer approximation
adopted in the literatures and find that it overestimates the $\gamma Z$ exchange effect significantly at the forward scattering angles.
\end{abstract}
\pacs{13.40.Ks, 13.60.Fz, 13.88.+e, 14.20.Dh}
\maketitle
The parity asymmetry
$A_{PV}=\sigma_R-\sigma_L/\sigma_R+\sigma_L$ in scattering of longitudinally polarized electrons elastically from unpolarized
protons is an important observable because it provides crucial information
of the strangeness content of the proton.
At the tree level,
the parity asymmetry $A_{PV}$ arises from the interference of diagrams
with one-photon-exchange (OPE) and $Z$-boson exchange shown in Fig. 1(a)
and (b), respectively.
\begin{figure}[t]
\centerline{\epsfxsize 1.4 truein\epsfbox{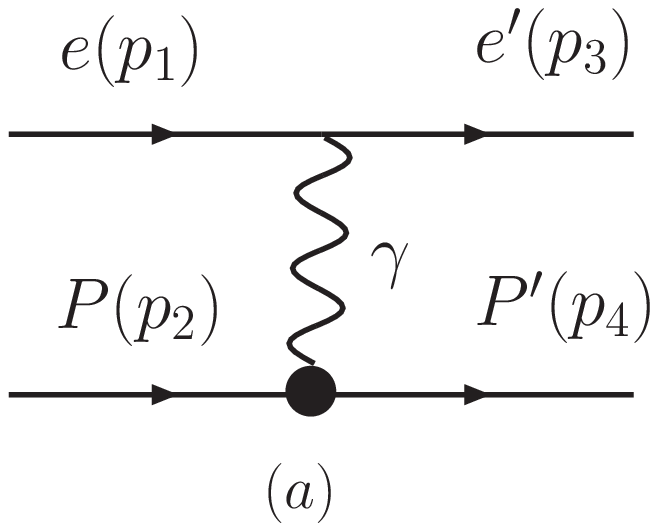}
\epsfxsize 1.4 truein\epsfbox{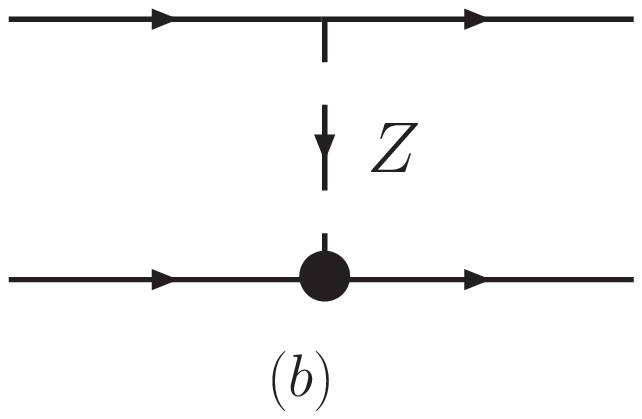}} \centerline{\epsfxsize
1.4 truein\epsfbox{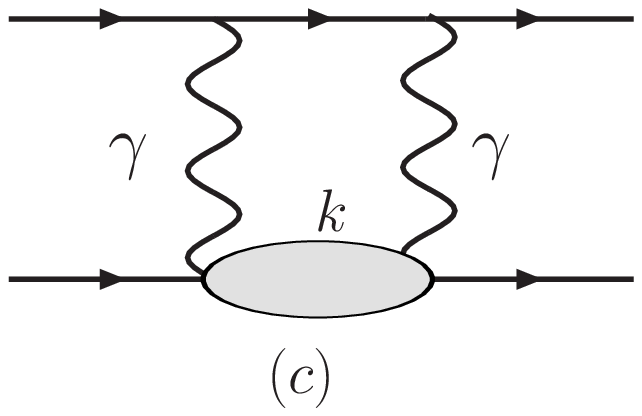} \epsfxsize 1.4
truein\epsfbox{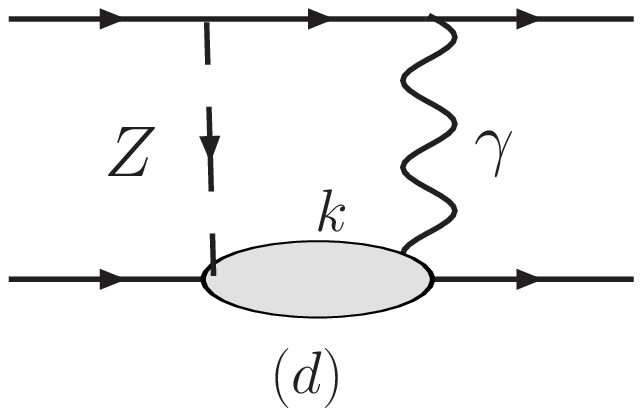}}
\caption{(a) one-photon-exchange,
(b) $Z$-boson-exchange, (c) TPE, and (d) $\gamma Z$ exchange
diagrams for elastic {\it ep} scattering. Corresponding cross-box
diagrams are implied.}\vskip -0.3cm
\end{figure}
The interactions between photon, $Z$-boson and proton
are described by the form factors of the proton defined as
\begin{eqnarray}
\langle p'|J^Z_\mu|p\rangle
&=&\overline{u}(p')\left[F^{Z p}_{1}\gamma_\mu+F^{Z p}_{2}\frac{i\sigma_{\mu\nu}q^\nu}{2M}
+G^{Zp}_A\gamma_\mu\gamma_5\right]u(p),\nonumber  \\ \nonumber
\langle p'|J^\gamma_\mu|p\rangle
&=&\overline{u}(p')\left[F^{\gamma p}_{1}\gamma_\mu+F^{\gamma p}_{2}
\frac{i\sigma_{\mu\nu}q^\nu}{2M}\right]u(p),
\end{eqnarray}
where $q=p'-p$ and $M$ is the mass of the proton.
$F^{\gamma p}_{1,2}, F^{Z p}_{1,2}$ and $G^{Zp}_{A}$
are the form factors of the proton electromagnetic and neutral weak current, respectively.
The parity asymmetry $A_{PV}$ can be expressed
by the form factors defined above as
\begin{equation}
A_{PV}(1\gamma+Z)
= -\frac{G_F
Q^2}{4\pi\alpha\sqrt{2}}\frac{A_E+A_M+A_A}{\bigl[\epsilon
(G^{\gamma p}_{E})^2+ \tau (G^{\gamma p}_{M})^2\bigr]},
\label{A_PV_Born}
\end{equation}
where $M$ is the proton mass and $G_{F}$ is Fermi constant and $Q^2$=$-q^2$, $\tau$=$Q^2/(4M^2)$, and $\epsilon=(1+2 (1+\tau)\tan^2\frac{\theta_e}{2})^{-1}$.
$A_{E}$, $A_{M}$ and $A_{A}$ are defined as
\begin{equation}
\begin{split}
&A_E=\epsilon G_{E}^{Z p} G_{E}^{\gamma p},
~~A_M=\tau G_{M}^{Z p} G_{M}^{\gamma p}, \\
&A_A=(-1+4\sin^2\theta_W)\sqrt{\tau (1+\tau) (1-\epsilon^2)} G_A^{Zp} G_{M}^{\gamma p},
\end{split}
\label{1gamma}
\end{equation}
where $\theta_W$ is the weak mixing (Weinberg) angle and
$G^{\gamma,Z p}_{E}=F^{\gamma,Z p}_{1}-\tau F^{\gamma,Z p}_{2}$,
$G^{\gamma,Z p}_{M}=F^{\gamma,Z p}_{1}+F^{\gamma,Z p}_{2}$.
When combined with proton
and neutron electromagnetic form factors and with the use of
charge symmetry, one obtains the following relation \cite{Kaplan}:
\begin{equation}
G^{Z p}_{E,M}=(1-4\sin^2\theta_W)G^{\gamma p}_{E,M}-G^{\gamma n}_{E,M}-G^s_{E,M}.
\label{strange}
\end{equation}
Consequently Eq.(\ref{A_PV_Born}) becomes
\begin{eqnarray}
&&A_{PV}(1\gamma+Z)=A_1+A_2+A_3, \nonumber\\
&& A_{1}= -a\left[1-4\sin^{2}\theta_{W}-\frac{\epsilon
G^{\gamma p}_{E}G^{\gamma n}_{E}
+\tau G^{\gamma p}_{M}G^{\gamma n}_{M}}{\epsilon(G^{\gamma p}_ {E})^2+\tau(G^{\gamma p}_{M})^2}\right],\nonumber \\
&& A_{2}= a\frac{\epsilon G^{\gamma p}_{E}G^{s}_{E}
+\tau G^{\gamma p}_{M}G^{s}_{M}}{\epsilon(G^{\gamma p}_{E})^2+\tau(G^{\gamma p}_{M})^2},\nonumber \\
&&A_{3}=a(1-4\sin^{2}\theta_{W})\frac{\sqrt{\tau (1+\tau) (1-\epsilon^2)}
G^{\gamma p}_{M}G_{A}^{Zp}}
{\epsilon(G^{\gamma p}_{E})^2+\tau(G^{\gamma p}_{M})^2},
\label{TreeA123}
\end{eqnarray}
where $a$=$\frac{G_F Q^2}{4\pi\alpha\sqrt{2}}$.
The strange form factors $G_{E,M}^{s}$ can be
determined through Eq.(\ref{TreeA123}).
Four experimental programs SAMPLE \cite{SAMPLE}, HAPPEX
\cite{HAPPEX}, A4 \cite{A4}, and G0 \cite{G0} have been designed
to measure $A_{PV}$, which is small and ranges from
0.1 to 100 ppm. Their results demonstrate that
the sum of $A_1$ and $A_3$ is much larger than $A_2$ and
the theoretical uncertainty is comparable in size to $A_2$.
This calls for greater
efforts to reduce the theoretical uncertainty in order to arrive at a
more precise determination of the strange form factors.

The main theoretical uncertainty is from higher-order electroweak
radiative corrections which have been carefully studied in
\cite{Marciano83,Musolf92}. Those corrections are from the interference
of the electroweak one-loop diagrams with the tree level diagrams.
Among those one-loop diagrams, the two-boson-exchange box diagrams shown
in Figs. 1(c) and 1(d) are particularly intricate
because they both depend on the nucleon structure.

Recently the Two-Photon-Exchange (TPE) diagram (Fig 1(c))
has been evaluated in \cite{Chen04}
in a parton model using GPDs where the handbag approximation is used, namely
the two photons are coupled directly to the same point-like quark.
The contribution of the interference of the
two-photon-exchange process of Fig. 1(c) with the one-boson-exchange diagrams
to $A_{PV}$, has been evaluated in \cite{Afanasev05}
in the same parton model used in \cite{Chen04}.
It was found that indeed the
TPE correction to $A_{PV}$ reaches a few percent and is non-negligible.

On the other hand,
the $\gamma Z$ exchange effects have been considered in
\cite{Marciano83} at quark and hadron levels, respectively.
They have however been computed
under a certain approximation where the momenta carried by the two
bosons were assumed to be equal, corresponding with the forward scattering case.
Furthermore the electron momenta were assumed to be zero, which is
relevant kinematical limit to calculate parity violation in atoms, considered in
\cite{Marciano83}.
To reduce the theoretical uncertainty at finite $Q^2$,
one must go beyond this particular approximation
in the calculation of the $\gamma Z$ exchange effects.

\begin{figure}
\includegraphics[width=3.0cm]{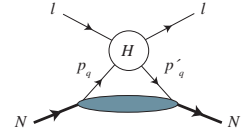}
\caption{Handbag approximation for the elastic lepton-nucleon scattering. In the partonic
process indicated by H, the lepton scatters from quarks within the nucleon, with momenta $P_{q}$ and $P_{q'}$.
The lower blob represents the
GPD's of the nucleon.}
\label{handbag}
\end{figure}

In this Letter, we report on the result of the partonic calculation of  $\gamma
Z$ exchange corrections to $A_{PV}$ in the
same parton model developed in \cite{Chen04}.
We first calculate the subprocess on a quark $e(k)+q(P_{q})\rightarrow e(k')+q(P_{q'})$,
denoted by the scattering amplitude $H$ in Fig. \ref{handbag}.
Subsequently we embed the quarks in the proton as described through the nucleon's
generalized parton distributions (GPDs). Here only GPDs of
$u$ and $d$ quarks are included because
the strange quark contribution to the box diagrams are
expected to be very small.

In the standard model the
electromagnetic current and the neutral weak current of quarks are
$J_{\mu}^{em}=\sum_{q} Q_{q}
\bar{q}\gamma_{\mu}q$ and
$J_{\mu}^{Z}=\sum_{q}\bar{q}(g^{q}_{V}+g^{q}_{A}\gamma_{5})q$.
Here we follow the notation of \cite{Musolf92} and have
$g^{u}_{V}=1-\frac{8}{3}\sin^{2}\theta_{W},\,\,g^{u}_{A}=-1$ and
$g^{d}_{V}=-1+\frac{4}{3}\sin^{2}\theta_{W},\,\,g^{d}_{A}=+1.$
The quark-level parity-violating amplitudes of the
$\gamma Z$ exchange diagrams are
\begin{eqnarray}
{\cal M}^{PV}_{\gamma Z}(eq\rightarrow eq)&=&
\frac{-i G_{F}}{2\sqrt{2}}
\sum_{q=u,d}
[t^{1}_{q}(\bar{u}_{e}\gamma_{\mu}\gamma_{5}u_{e})
(\bar{q}\gamma^{\mu}q) \nonumber \\
&+&t^{2}_{q}(\bar{u}_{e}\gamma_{\mu}u_{e})(\bar{q}\gamma^{\mu}\gamma_{5}q)],
\label{ourM}
\end{eqnarray}
where $t^{q}_{1}$=$Q_{q}[c_{1}g^{e}_{A}g^{q}_{V}+c_{2}g^{q}_{A}g^{e}_{V}]$ and
$t^{q}_{2}$=$Q_{q}[c_{1}g^{e}_{V}g^{q}_{A}+c_{2}g^{q}_{V}g^{e}_{A}]$.
Here $g^{e}_{V}$=$-1+4\sin^{2}\theta_{W}$, $g^{e}_{A}$=$+1$.
$c_{1}$ and $c_{2}$ are defined as
\begin{eqnarray}
&&c_{1}=\frac{-e^2}{4\pi^2}\left[\ln\left(\frac{\lambda^2}{Q^2}\right)+\frac{\pi^2}{2}\right]+
\frac{3e^2}{16\pi^2}\ln\left(\frac{\hat{u}}{\hat{s}}\right),\nonumber \\
&&c_{2}=\frac{e^2}{16\pi^2}\left[-7+3\ln\left(\frac{\hat{s}}{M_{Z}^{2}}\right)
+3\ln\left(\frac{\hat{u}}{M_{Z}^{2}}\right)\right],
\end{eqnarray}
where $\hat{s}=(P_{q}+k)^2,\hat{u}=(P_q-k')^2$ and $\lambda$ is the infrared cut-off
input by infinitesimal photon mass.
The amplitudes are separated into the soft and hard parts, the soft part corresponds with the
situation where the photon carries zero four momentum and one obtains
$c_{1}^{soft}=\frac{-e^2}{4\pi^2}\left[\ln\left(\frac{\lambda^2}{Q^2}\right)+\frac{\pi^2}{2}\right]$,
the hard part is
$c_{1}^{hard}=\frac{3e^2}{16\pi^2}\ln\left(\frac{\hat{u}}{\hat{s}}\right)$.
The IR divergence arising from the direct and crossed box diagrams is concealed
with the bremsstrahlung interference contribution with a soft photon emitted from the
electron and the proton. We limit ourselves to discuss the hard $\gamma Z$ exchange contribution
in this article.
The $ep$ parity-violating amplitudes are
obtained as a convolution between an electron-quark hard scattering and a soft nucleon matrix element.
The procedure is similar to the one in \cite{Chen04}.
The parity-violating $ep$ amplitude can be written as
\begin{eqnarray}
{\cal M}^{PV,Parton}_{\gamma Z}&=&-i\frac{G_{F}}{2\sqrt{2}}[\bar{u}_{e}\gamma_{\mu}u_{e}
\bar{u}_{N}[ \Gamma_{1}\gamma^{\mu}]\gamma_{5}u_{N}\nonumber \\
&+&\bar{u}_{e}\gamma_{\mu}\gamma_{5}u_{e}
\bar{u}_{N}
\left[\Gamma_{2}\gamma^{\mu}-\Gamma_{3}\frac{P^{\mu}}{M}\right]u_{N}].
\end{eqnarray}
$\Gamma_{1}$, $\Gamma_{2}$ and $\Gamma_{3}$ are defined as
\begin{eqnarray}
\Gamma_{1}&=& \frac{1+\epsilon}{2\epsilon}F-\frac{1+\epsilon}{2\epsilon}\frac{Q^2}{s-u}D, \nonumber \\
\Gamma_{2}&=&\frac{1+\epsilon}{2\epsilon}D-\frac{1+\epsilon}{2\epsilon}\frac{Q^2+4M^2}{s-u}F, \nonumber \\
\Gamma_{3}&=& \frac{1}{1+\tau}\left[\Gamma_{2}-\sqrt{\frac{1+\epsilon}{2\epsilon}}E\right],
\end{eqnarray}
where $D$, $E$ and $F$ are defined as
\begin{eqnarray}
D&\equiv&\int^{1}_{-1}\sum_{q=u,d}\frac{dx}{x}\frac{Q^2 t^{q}_{1}+(\hat{s}-\hat{u})t^{q}_{2}}{s-u}(H^{q}+E^{q}),\nonumber \\
E&\equiv&\int^{1}_{-1}\sum_{q=u,d}\frac{dx}{x}\frac{Q^2 t^{q}_{1}+(\hat{s}-\hat{u})t^{q}_{2}}
{s-u}(H^{q}-\tau E^{q}),\nonumber \\
F&\equiv&\int^{1}_{-1}\sum_{q=u,d}\frac{dx}{x}\frac{Q^2 t^{q}_{2}+(\hat{s}-\hat{u})t^{q}_{1}}
{s-u}sgm(x)\tilde{H}^{q},
\label{DEF}
\end{eqnarray}
where $s$=$(p_1+p_3)^2$, $u$=$(p_2-p_3)^2$ and
$H^{q}$, $E^{q}$ and $\tilde{H}^{q}$ are the GPDs for a quark in the nucleon.
To estimate the amplitudes of Eq. (\ref{DEF}) one needs to specify a model for the GPDs.
Again we follow \cite{Chen04} to adopt an unfactorized model of GPDs in terms of
a forward parton distributions and a gaussian factor in $x$ and $-Q^2$.
For the details we refer the readers to \cite{Chen04}.
\begin{figure}
\includegraphics[width=8.0cm]{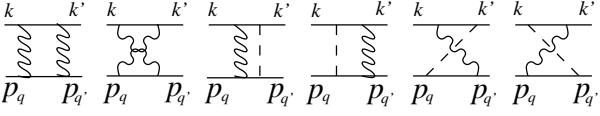}
\vskip -0.3cm
\caption{The direct and crossed box diagrams to describe the TPE and $\gamma Z$ exchange contribution to the lepton-quark scattering, corresponding with the blob denoted by $H$ in Fig. 2.}
\label{direct}\vskip -0.3cm
\end{figure}
\begin{figure}[t]
\centerline{\epsfxsize 1.4 truein\epsfbox{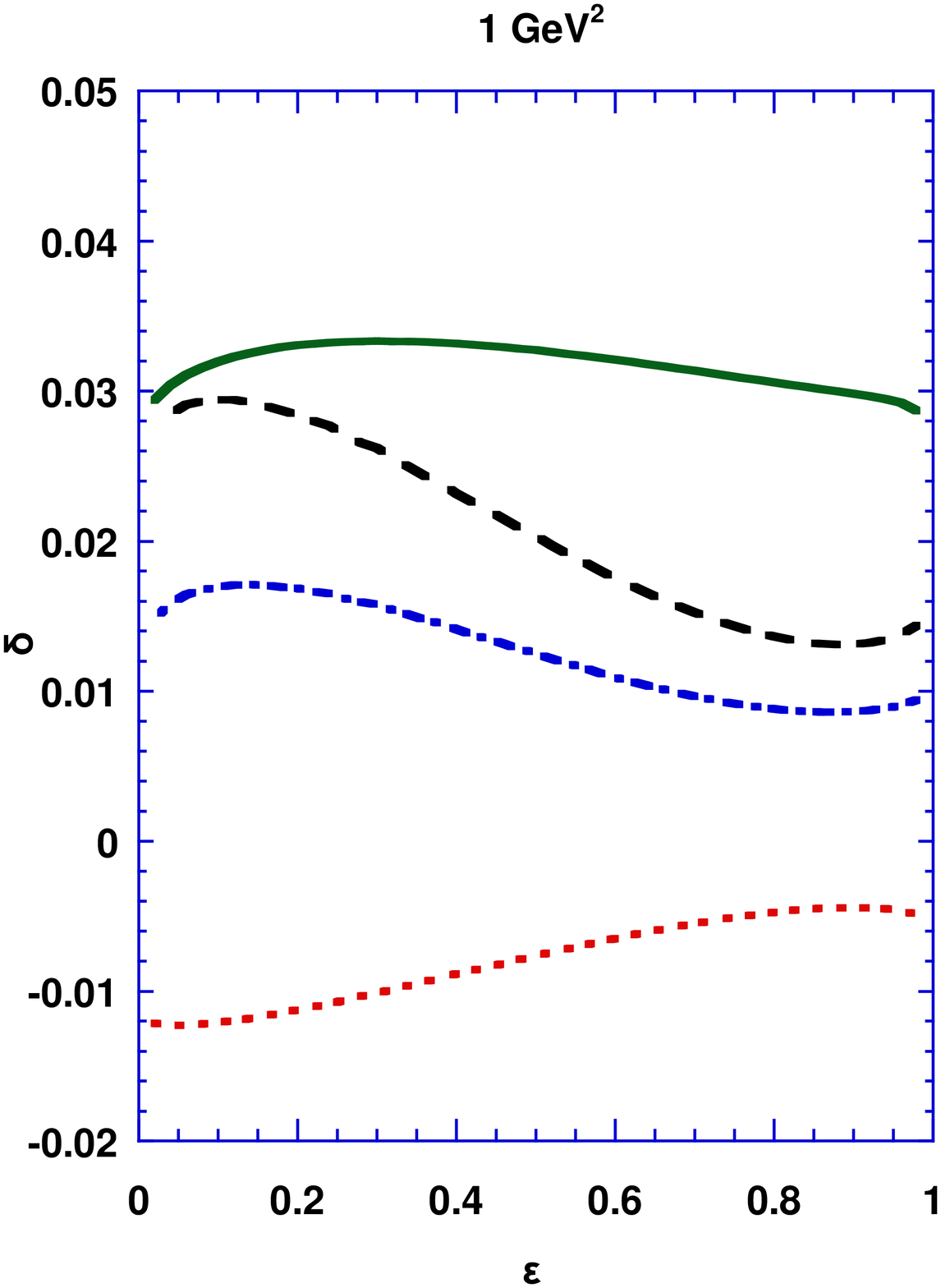}
\epsfxsize 1.4 truein\epsfbox{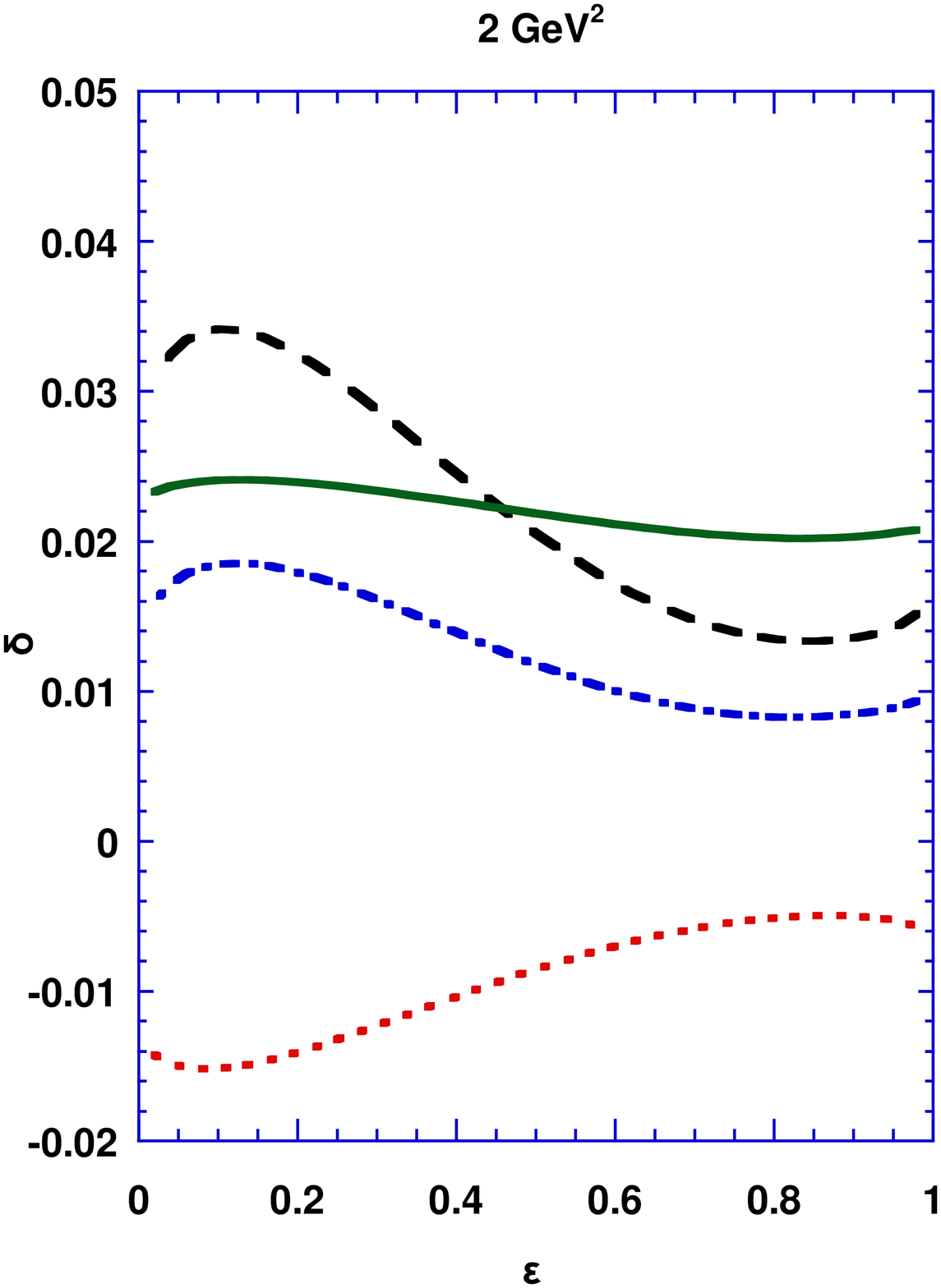}} \centerline{\epsfxsize
1.4truein\epsfbox{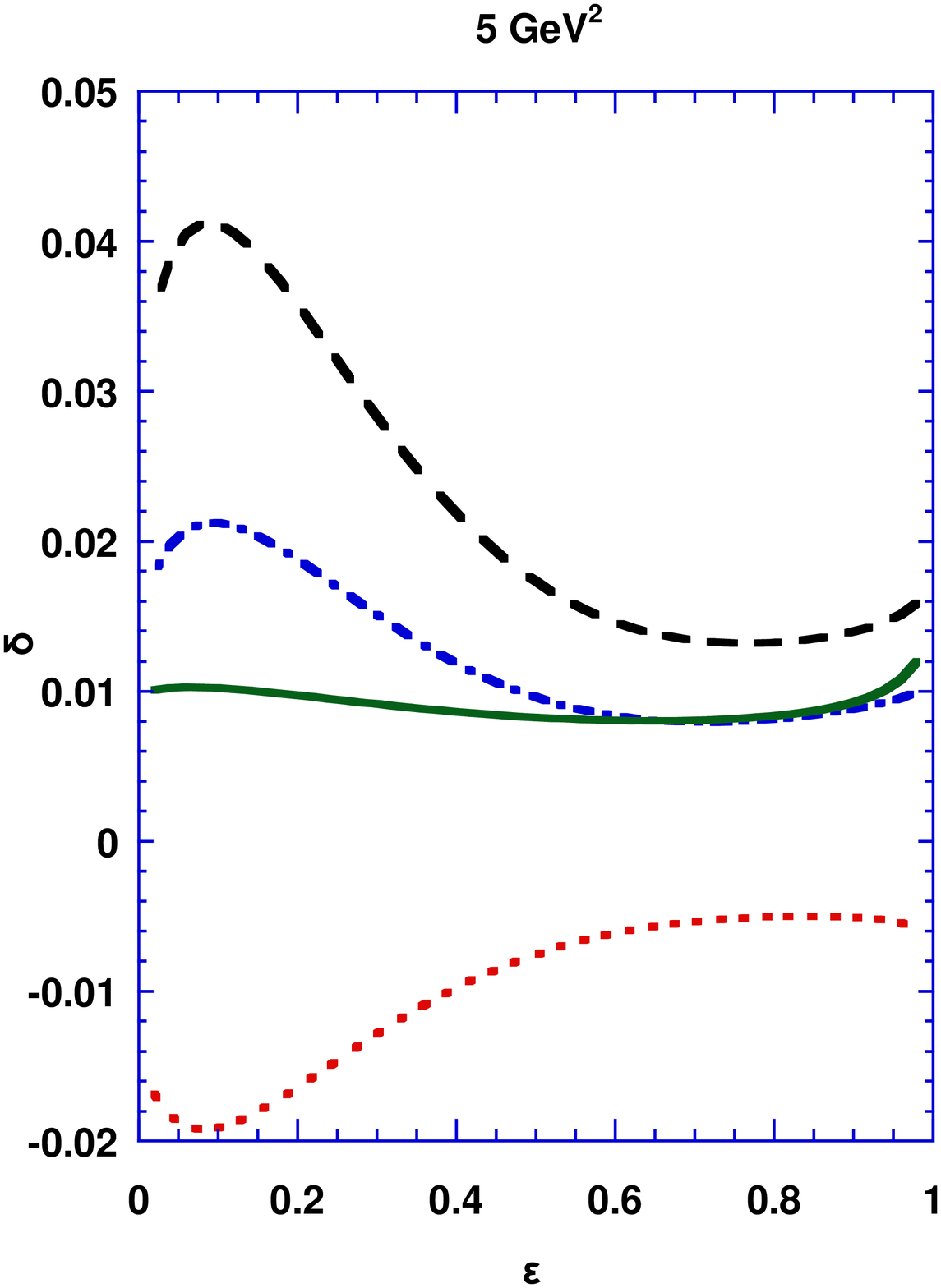} \epsfxsize 1.4
truein\epsfbox{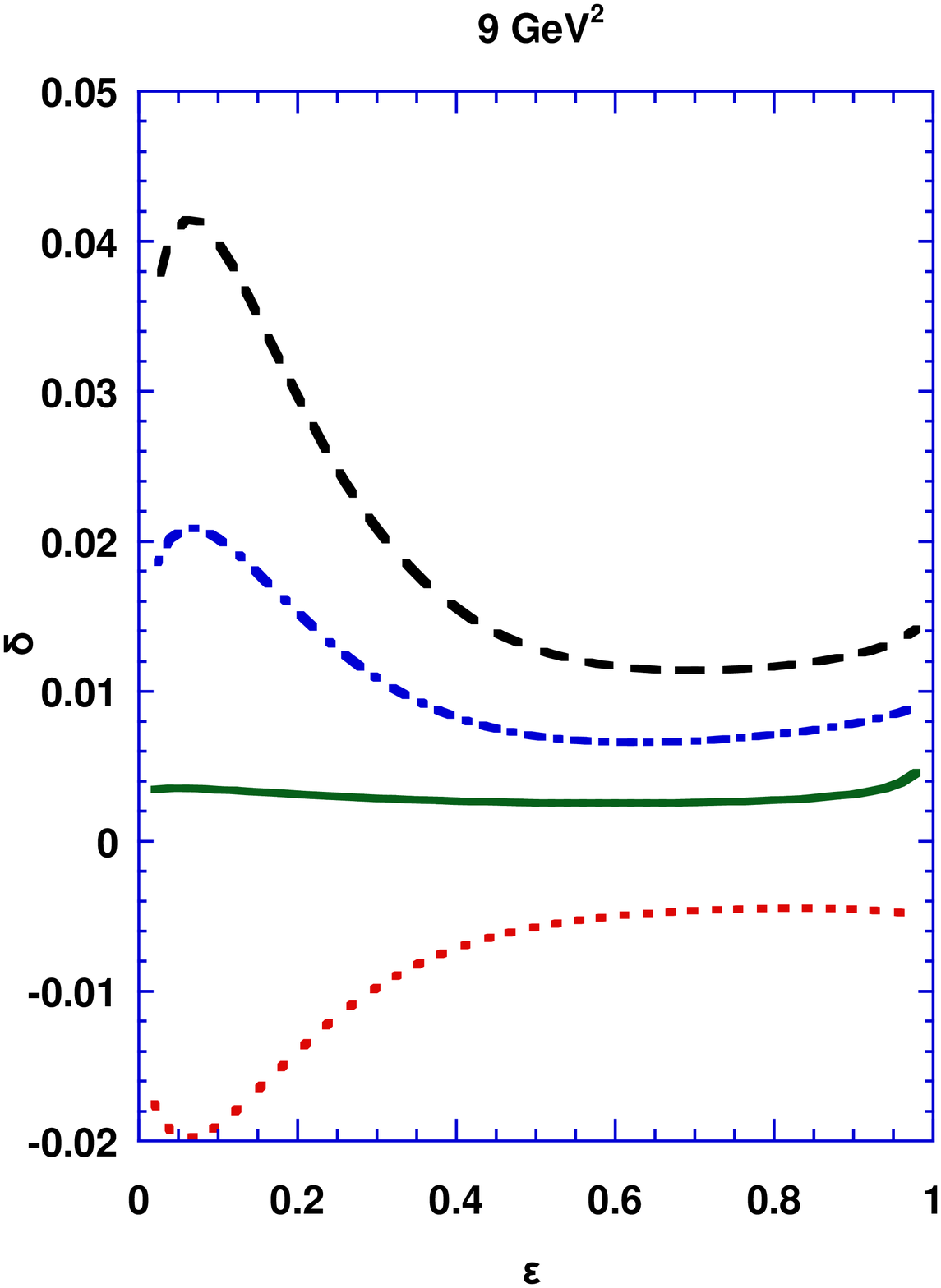}}
\caption{TPE and $\gamma Z$ exchange corrections to parity-violating
asymmetry as functions of $\epsilon$.
The left of upper panel for $Q^2$=1 GeV$^2$, the
right for $Q^2$=2 GeV$^2$. The left of lower panel for $Q^2$=5 GeV$^2$,
the right for $Q^2$=9 GeV$^2$. The various contributions are $\gamma(\gamma\gamma)$(dashed line),
$Z(\gamma\gamma)$(dotted lines), TPE total contribution $\delta_{2\gamma}$(dash-dotted lines)
and $\delta_{\gamma Z}$ (solid lines).
}
\label{Mainresult}\vskip -0.1cm
\end{figure}
In Fig. \ref{Mainresult}, we show the TPE and $\gamma Z$ exchange corrections to
$A_{PV}$ by plotting $\delta$, defined by
\begin{equation}A_{PV}(1\gamma+Z+2\gamma+\gamma
Z)=A_{PV}(1\gamma+Z)(1+\delta_{2\gamma}+\delta_{\gamma Z}),\nonumber
\end{equation}
at four
different values of $Q^2$=1, 2, 5 and 9 GeV$^2$.
$A_{PV}(1\gamma+Z+2\gamma+\gamma Z)$
includes the effects of TPE and $\gamma Z$ exchange.
The interferences between OPE and $\gamma Z$ exchange denoted by
$\delta_{\gamma Z}$ are represented by solid lines.
The effects of the parity conserving interference between OPE and TPE
$(\gamma (\gamma \gamma))$,
entering the denominator of the asymmetry, are represented by dashed lines.
The interference between $Z$-exchange and TPE ($Z(\gamma\gamma)$) are represented by dotted
lines, with their sum ($\delta_{2\gamma}$) denoted by dot-dashed lines. Our result of
$\gamma(\gamma\gamma)$ and $Z(\gamma\gamma)$ is in agreement with \cite{Afanasev05}.

Our result shows that TBE effects are about few percent as expected. However
$\delta_{\gamma Z}$ and $\delta_{2\gamma}$ behave quite differently.
In handbag calculation,
$\delta_{\gamma Z}$ is sensitive to $Q^2$. For example, its value is tripled
when $Q^2$ decreases from 5 GeV$^2$ to 1 GeV$^2$. On the other hand, the value
of $\delta_{2\gamma}$ is between $1\sim 2\%$ at all $Q^2$ values.
Furthermore
the value of $\delta_{2\gamma}$ at backward angles is in general lager than the one at
forward angles but $\delta_{\gamma Z}$ is insensitive to $\epsilon$.
On the contrary, the result of the hadronic model \cite{Zhou07} shows that both of
$\delta_{\gamma Z}$ and $\delta_{2\gamma}$ have very strong $\epsilon$ dependencies and
always decrease into zero when $\epsilon$ approaches one. Such a difference may be
able to be explained by the following:
In the hadronic model the loop
momentum integration is mostly from the low loop momenta
because the form factors are inserted as regulators. However, in the partonic calculation
the loop momentum integration is dominated by the high loop momentum.
Naively one should add the results of the two calculation together.
How to combine the results of these two calculations remains an open issue.
\begin{figure}[hbtp]
\includegraphics[width=3.6cm]{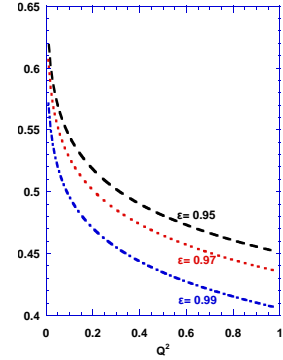}
\caption{$\Delta_{MS}$ as functions of $Q^2$ (ranging from 0.03 to 1 GeV$^2$) at $\epsilon=0.95$ (dashed line), $\epsilon=0.97$ (dotted line)
and $\epsilon=0.99$ (solid line).}
\label{MS}\vskip -0.2cm
\end{figure}
Another important issue to address is the validity of the previous
estimate of $\gamma Z$ exchange effects made by Marciano and Sirlin
\cite{Marciano83}
because their estimate is widely used.
They evaluated the $\gamma Z$ box diagrams in Fig. 3 in the limit of
$k=k'=0$ and $P_{q}=P_{q'}$. Their result shows that
\begin{eqnarray}
&&{\cal M}^{PV,MS}_{\gamma Z}(eq\rightarrow eq)=
\frac{-i G_{F}}{\sqrt{2}}
\sum_{q=u,d}
[C_{1q}\nonumber \\
&&(\bar{u}_{e}\gamma_{\mu}\gamma_{5}u_{e})
(\bar{q}\gamma^{\mu}q)
+C_{2q}(\bar{u}_{e}\gamma_{\mu}u_{e})(\bar{q}\gamma^{\mu}\gamma_{5}q)],\nonumber \\
\label{MsM}
\end{eqnarray}
here $C_{1q}$ and $C_{2q}$ are
\begin{eqnarray}
&&C_{1u}=\frac{\alpha}{2\pi}(1-4\sin^2\theta_{W}){\cal C},\,\,
C_{1d}=\frac{\alpha}{4\pi}(1-4\sin^2\theta_{W}){\cal C},\nonumber \\
&&C_{2u}=\frac{\alpha}{2\pi}(1-\frac{8}{3}\sin^2\theta_{W}){\cal C},\,\,
C_{2d}=\frac{\alpha}{4\pi}(1-\frac{4}{3}\sin^2\theta_{W}){\cal C},
\nonumber
\end{eqnarray}
here ${\cal C}=\ln\frac{M_Z^2}{M^2_{A}}+\frac{3}{2}$. $M_{A}$ is a hadronic mass
scale associated with the asymptotic behavior and its value is set to be 1 GeV \cite{Marciano83}.
Because $C_{1q}$ and $C_{2q}$ are constants,
the amplitude of Fig 1(d) can be written as
\begin{eqnarray}
&&{\cal M}^{PV,MS}_{\gamma Z}={\cal M}^{PV,MS}_{\gamma Z, V}+{\cal M}^{PV,MS}_{\gamma Z,A},\nonumber \\
&&{\cal M}^{PV,MS}_{\gamma Z, V}=\frac{-i G_{F}}{\sqrt{2}}\bar{u}_{e}\gamma_{\mu}\gamma_{5}u_{e} \sum_{q=u,d}C_{1q}
\bar{u}_{N}[\bar{q}\gamma^{\mu}q]u_{N}, \nonumber \\
&&{\cal M}^{PV,MS}_{\gamma Z, A}=\frac{-i G_{F}}{\sqrt{2}} \bar{u}_{e}\gamma_{\mu}u_{e}\sum_{q=u,d}C_{2q}
\bar{u}_{N}[\bar{q}\gamma^{\mu}\gamma_{5}q]u_{N}. \nonumber \\
\nonumber
\end{eqnarray}
\vskip -0.1cm
According to charge symmetry, the form factors associated with the matrix elements of
quark vector current can be written as
\begin{eqnarray}
G^{u/p}_{E,M}&=&2G^{\gamma p}_{E,M}+G^{\gamma n}_{E,M}+G^{s}_{E,M}, \nonumber \\
G^{d/p}_{E,M}&=&G^{\gamma p}_{E,M}+2G^{\gamma n}_{E,M}+G^{s}_{E,M}.
\end{eqnarray}
\vskip -0.1cm
If we follow notation from \cite{Marciano83}:
\begin{eqnarray}
C_{1u}&=&\frac{1}{2}\Delta\rho\left[1-\frac{8}{3}\Delta\kappa \sin^{2}\theta_{W}\right],\nonumber \\
C_{1d}&=&\frac{-1}{2}\Delta\rho\left[1-\frac{4}{3}\Delta\kappa \sin^{2}\theta_{W}\right],
\end{eqnarray}
then one obtains the following widely used formula,
\begin{eqnarray}
&&A^{MS}_{PV}(\gamma Z)=\frac{2Re[{\cal M}_{1\gamma}^{\dagger}{\cal M}^{PV,MS}_{\gamma Z}]}{|{\cal M}_{1\gamma}|^{2}}=\tilde{A}_1+\tilde{A}_2+\tilde{A}_3, \nonumber\\
&& \tilde{A}_{1}=-a\Delta\rho\left[1-4\Delta\kappa\sin^{2}\theta_{W}-\frac{\epsilon
G^{\gamma p}_{E}G^{\gamma n}_{E}
+\tau G^{\gamma p}_{M}G^{\gamma n}_{M}}{\epsilon(G^{\gamma p}_ {E})^2+\tau(G^{\gamma p}_{M})^2}\right],\nonumber \\
&& \tilde{A}_{2}= a\Delta\rho\frac{\epsilon G^{\gamma p}_{E}G^{s}_{E}
+\tau G^{\gamma p}_{M}G^{s}_{M}}{\epsilon(G^{\gamma p}_{E})^2+\tau(G^{\gamma p}_{M})^2},\nonumber \\
&&\tilde{A}_{3}=a(1-4\sin^{2}\theta_{W})\frac{\sqrt{\tau (1+\tau) (1-\epsilon^2)}
G^{\gamma p}_{M}\tilde{G}_{A}^{Z}}
{\epsilon(G^{\gamma p}_{E})^2+\tau(G^{\gamma p}_{M})^2},
\label{MS2}
\end{eqnarray}
here $\tilde{G}_{A}$ absorbs the contribution from ${\cal M}^{PV,MS}_{\gamma Z,A}$ which is suppressed
at the forward scattering angles.

We therefore define the following quantity to characterize the difference between
our partonic calculation and the previous estimate taken at the limit $k=k'=0$ and $P_{q}=P_{q'}$:
\begin{equation}
\Delta_{MS}=\frac{A_{PV}^{Parton}(\gamma Z)}{A_{PV}^{MS}(\gamma Z)}=
\frac{Re[{\cal M}_{1\gamma}^{\dagger}{\cal M}^{PV,Parton}_{\gamma Z}]}
{Re[{\cal M}_{1\gamma}^{\dagger}{\cal M}^{PV,MS}_{\gamma Z}]}. \nonumber
\end{equation}
If MS's approximation overestimates the $\gamma Z$ exchange effect then $\Delta_{MS}$ will be smaller than one.
$\Delta_{MS}$ is a function of $\epsilon$ and $Q^2$. Here we are mostly concerned about the case of low $Q^2$ and
the forward angles.
In Fig. \ref{MS} we choose $\epsilon=0.99$ (dash-dotted line), 0.97(dotted line) and
0.95(dashed line) and 0.03 GeV$^2 \le Q^2 \le$ 1 GeV$^2$. It shows that MS's approximation
always overestimates the $\gamma Z$ exchange effect and the extent of overestimate grows as $Q^2$ and $\epsilon$
increase. Note that $\Delta_{MS}$ grows
fast when $Q^2\le$ 0.1 GeV$^2$ which means $A_{PV}^{Parton}(\gamma Z)$ grows fast there. It is interesting because
similar feature also appears in the result of hadronic model \cite{Zhou07}.
However even when $Q^2$=0.03 GeV$^2$ and $\epsilon$=0.99, $\Delta_{MS}\approx 0.55$ which means MS's approximation still overestimates $\gamma Z$ exchange effect up to $45\%$.
It is obvious that $A_{PV}^{MS}(\gamma Z)$ is significantly larger than $A_{PV}^{Parton}(\gamma Z)$.
Hence we corroborate the claim made by \cite{Zhou07} that the $\gamma Z$ exchange effect is indeed overestimated by the zero-momentum-transfer approximation adopted by the previous estimate.

We conclude that the
TPE and $\gamma Z$ exchange effects contribute to the parity asymmetry at a level of
a few percent according to our partonic calculation.
In particular, the $\gamma Z$ exchange effect is sensitive to
$Q^2$ and its magnitude increases when $Q^2$ decreases but it is insensitive to $\epsilon$.
We also compare the $A_{PV}(\gamma Z)$ under the approximation relevant for the calculation of atomic
parity violation with
$A_{PV}(\gamma Z)$ based on our partonic calculation. We find that $A_{PV}^{MS}(\gamma Z)$ is
significantly larger than $A_{PV}^{Parton}(\gamma Z)$.
Therefore we corroborate the claim made by \cite{Zhou07} that $\gamma$-$Z$ exchange effect is indeed overestimated by the zero momentum-transfer approximation.
\vskip -0.1cm
The authors would like to thank C. Carlson for useful discussions.
This work is partially supported by the National Science Council
of Taiwan under grants nos. NSC96-2112-M033-003-MY3 (C.W.K.) and
NSC96-2811-M-001-087 (Y.C.C.).

\end{document}